\RequirePackage[2020-02-02]{latexrelease}
\documentclass[aps,prb,twocolumn]{revtex4}
\usepackage{ucs}
\usepackage{graphicx}
\input epsf
\usepackage{amsmath}
\usepackage{amssymb}
\usepackage{amsfonts}
\usepackage{mathrsfs}
\usepackage{float}
\usepackage{color,soul}

\begin{document}
\title{Superinsulating behavior in granular Pb film on gated few-layer MoS$_2$}
\author{Suraina Gupta}
\affiliation{Department of Physics, Indian Institute of Technology Kanpur, Kanpur 208016, India}
\author{Santu Prasad Jana}
\affiliation{Department of Physics, Indian Institute of Technology Kanpur, Kanpur 208016, India}
\author{Pawan Kumar Gupta}
\affiliation{Department of Physics, Indian Institute of Technology Kanpur, Kanpur 208016, India}
\author{Anjan K. Gupta}
\affiliation{Department of Physics, Indian Institute of Technology Kanpur, Kanpur 208016, India}

\date{\today}

\begin{abstract}
We report a super-insulating behavior, in a device having granular Pb film on back-gated few-layer $\mathrm{MoS_2}$, below an onset temperature same as the critical temperature $T_{\rm C}\approx7$ K of bulk Pb. Below $T_{\rm C}$, the current-voltage characteristics exhibit a threshold voltage marking a crossover between the low-bias insulating and the high-bias normal-resistance states, consistent with the known super-insulating state behavior. A temperature dependent critical magnetic field is also found above which the insulating behavior is suppressed. The threshold voltage is found to vary with the gate-voltage but the critical field remains unchanged. With reducing temperature, the sample conductance saturates to a finite value, which depends on magnetic field and gate-voltage. This saturation behavior is found to be inconsistent with the charge-BKT and the thermal activation models but it can be fitted well to a combination of thermal activation and quantum fluctuations.
\end{abstract}
\maketitle

\section{Introduction}

The superconductor-to-insulator quantum phase transition (SIT) \cite{sondhi1997continuous, fisher1990quantum} has been of significant research focus over past few decades. There are several intriguing phenomena associated with this transition that include the emergence of a Bose metal phase at low temperatures \cite{tsen2016nature, sharma20182d}, scaling behavior near the quantum critical point \cite{allain2012electrical, breznay2017superconductor}, quantum Griffiths singularity \cite{xing2015quantum, chen2024quantum}, a change in the sign of magnetoresistance along with giant peaks \cite{nguyen2009observation, hollen2014fate}, and thermally activated behavior in the Cooper pair (CP) insulator phase \cite{baturina2007localized, stewart2007superconducting}. The SIT arises due to a competition between the transport of vortices and CPs. Thus, the complementarity between the CP number $N_{\rm s}$ and the phase $\phi$ of the condensate, leading to the uncertainty relation $\Delta \phi\:\Delta N_{\rm s} \geq 1$, plays an important role in the SIT physics.

Two-dimensional (2D) granular superconductors \cite{khan2000superconductor, hen2021superconductor} and Josephson junction (JJ) arrays \cite{van1996quantum, han2014collapse} are two model systems that exhibit SIT. In these systems, the Coulomb charging energy $E_{\rm C}$, promoting the localized CPs, favors a well-defined $N_{\rm s}$ state, while the Josephson coupling energy $E_{\rm J}$, promoting inter-grain phase coherence, favors a well-defined $\phi$ state. A zero resistance superconducting (SC) state occurs when $E_{\rm J} > E_{\rm C}$ and CPs form a coherent condensate while vortices are localized. On the other hand, for $E_{\rm C} > E_{\rm J}$, CPs become localized and vortices move freely. This leads to a `zero conductivity' superinsulating (SI) state, at very low temperatures, which is considered dual to the SC state \cite{vinokur2008superinsulator, fistul2008collective, baturina2013superinsulator}. The duality reflects a role reversal between CPs and vortices, or between $\phi$ and $N_{\rm s}$, as well as between `current and voltage' in the two extreme dissipationless states.

SIT occurs when a non-thermal parameter is varied which changes the ratio $E_{\rm J}/E_{\rm C}$. Experimentally, this is achieved by varying the film thickness, applying magnetic fields, or using a gate voltage to tune the carrier density. In the `Bosonic' route to SIT, the transition arises from the localization of CPs \cite{baturina2007localized, steiner2008approach, hollen2011cooper}. In contrast, the `fermionic' pathway involves breaking of CPs and suppressing the SC order parameter amplitude \cite{valles1992electron, hsu1993magnetic, hollen2013collapse}. However, the actual such systems often display features from both scenarios, making their behavior more intricate than either of the two models.
\begin{figure}[h]
	\centering
 	\includegraphics[width=3.2in]{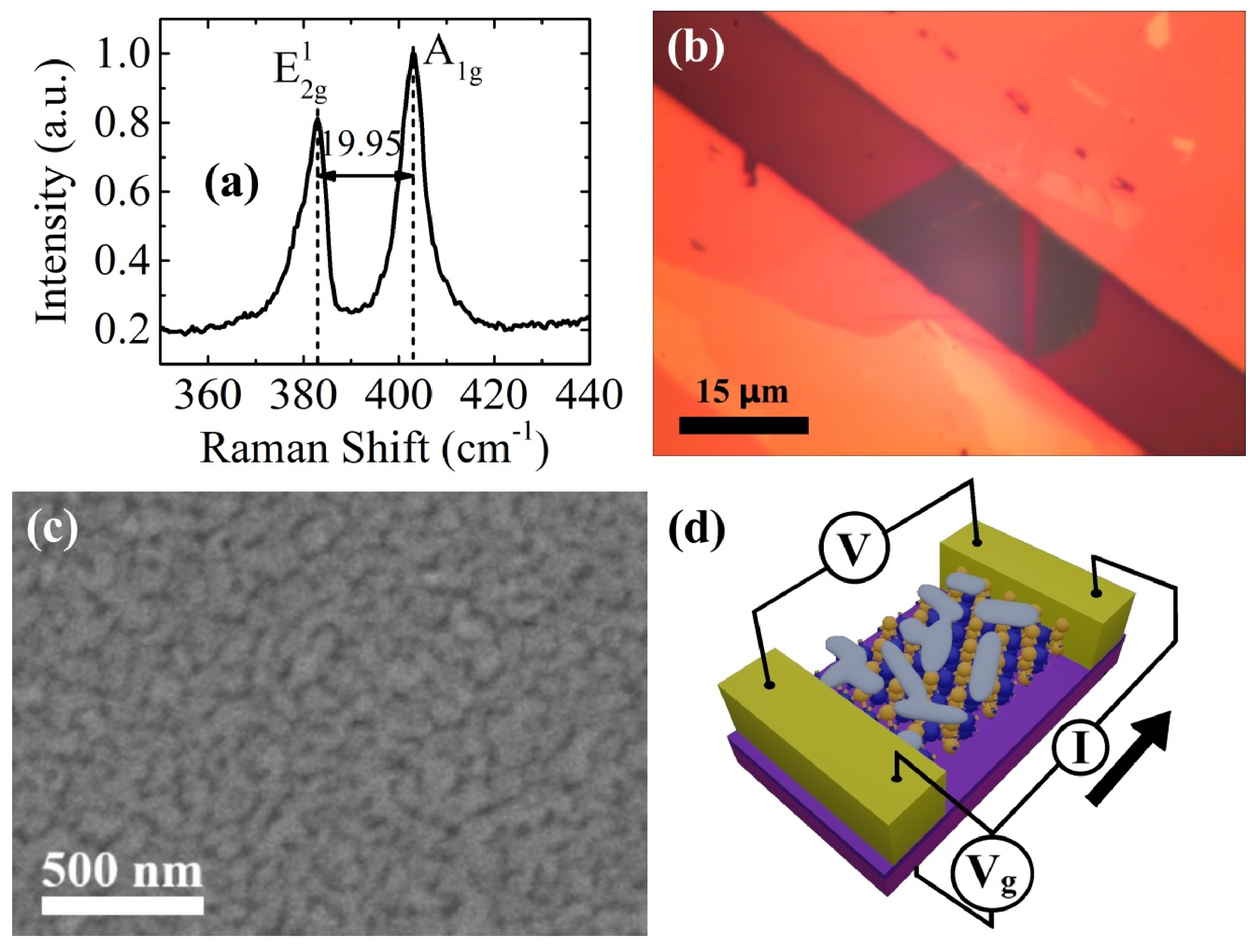}
	\caption{(a) shows the Raman spectrum of the exfoliated $\mathrm{MoS_2}$ with two characteristic Raman peaks at 382.94 $\mathrm{cm^{-1}}$ ($\mathrm{E_{2g}^1}$) and 402.89 $\mathrm{cm^{-1}}$ ($\mathrm{A_{1g}}$). The 19.95 $\mathrm{cm^{-1}}$ peak separation confirms it to be few-layer $\mathrm{MoS_2}$. (b) Optical image of $\mathrm{MoS_2}$ (dark contrast) with gold-film contacts. (c) An SEM image showing islands of Pb (grey color) on $\mathrm{MoS_2}$ (black background). (d) shows the two probe transport configuration with bias current $I$, sample voltage $V$ and back-gate voltage $V_{\rm g}$.}
	\label{fig:INS1}
\end{figure}

The characteristic features of such insulating phases, such as thermally activated conductivity and threshold voltage, have been observed in materials like amorphous indium oxide \cite{sambandamurthy2004superconductivity, sambandamurthy2005experimental} and titanium nitride \cite{baturina2007localized} thin films. These experimental observations were theoretically addressed by Vinokur and co-workers \cite{vinokur2008superinsulator, fistul2008collective, baturina2013superinsulator} and others \cite{beloborodov2007granular, syzranov2009dc}. Hybrid devices consisting of superconductors and 2D materials have emerged recently as promising platforms for SIT studies. Allain et al. \cite{allain2012electrical}, using tin islands on graphene, demonstrated the tunability of the Josephson coupling through back-gate, enabling gate-induced SIT. Semiconducting molybdenum disulfide $(\mathrm{MoS_2})$, with its large bandgap \cite{radisavljevic2011single, yoon2011good}, offers an alternative to graphene. In particular, this enables a weak $E_{\rm J}$ regime for investigating SI states and other quantum phenomena.

This paper presents a temperature-dependent transport study on a hybrid system of granular lead (Pb) film on exfoliated few-layer $\mathrm{MoS_2}$, as a function of back-gate voltage and magnetic field. An increase in resistance is seen below the critical temperature $T_{\rm C}$ of bulk Pb. This insulating behavior, correlated with the superconductivity in Pb islands, is observed below a threshold sample voltage and for magnetic fields below a critical value. The conductivity saturates to a non-zero value for low temperatures down to 1.3 K. This is successfully modeled using a combination of thermal activation and quantum fluctuations. Thus, the gated MoS$_2$–Pb hybrid system provides another platform for studying such correlated quantum phases.

\section{Experimental Details}
Highly p-doped silicon wafers with a 300 nm oxide coating were first cleaned for 5 minutes by sonication in acetone and isopropyl alcohol, each. Next, few-layer $\mathrm{MoS_2}$ flakes were mechanically exfoliated from natural bulk crystal (from SPI) and were stamped on the cleaned $\mathrm{SiO_2}$/Si substrates, using a conventional dry transfer method \cite{castellanos2014deterministic}. This process employed polydimethylsiloxane (PDMS) films (from Gel-Pak) as a viscoelastic stamp, and an XYZ micro-manipulator attached to an optical microscope. The number of $\mathrm{MoS_2}$ layers on the substrate was then determined using Raman Spectroscopy. From the Raman spectrum, see Fig. \ref{fig:INS1}(a), the peak frequency separation between $\mathrm{E_{2g}^1}$ and $\mathrm{A_{1g}}$ is found as 19.95 cm$^{-1}$, which corresponds to the few-layer $\mathrm{MoS_2}$ \cite{chakraborty2012symmetry,lee2010anomalous}.

Two-probe gold-film (50 nm) contacts were made on $\mathrm{MoS_2}$ using mechanical wire-masking to avoid contamination from lithography resists and chemicals. Fig. \ref{fig:INS1}(b) shows an optical image of the studied $\mathrm{MoS_2}$ device with Au contacts. Subsequently, 40 nm Pb film was thermally evaporated on this sample surface at 10 \AA/sec rate with the substrate kept at room temperature. The high deposition rate ensures a high-quality interface between Pb and $\mathrm{MoS_2}$. As seen in the scanning electron micrograph in Fig. \ref{fig:INS1}(c), Pb forms a granular film on $\mathrm{MoS_2}$ due to its poor wettability. Some portion of the $\mathrm{SiO_2}$ substrate surrounding $\mathrm{MoS_2}$ was also deposited with Pb, but a large and clear separation between Pb islands was found on $\mathrm{SiO_2}$ as compared to $\mathrm{MoS_2}$. Thus, electrical conduction occurs only through Pb islands on $\mathrm{MoS_2}$.

After fabrication, the sample was promptly mounted on a cryostat, which was then cooled in a closed-cycle helium refrigerator to its base temperature of 1.3 K after evacuation. Low-pass RC-filters with a 15 kHz cutoff frequency, along with high-frequency-cutoff pi-filters, were integrated into the measurement lines at room temperature to reduce the electromagnetic noise. The transport measurement wires were also routed through Cu-powder filters at the base temperature to further alleviate the noise interference. The resistance measurements were carried out by measuring the sample voltage at small bias currents of either 10 or 100 nA, see the inset of Fig. \ref{fig:INS1}(d). The voltage-current characteristics (VICs) were acquired by applying current between $-0.5$ and $0.5\;\mu$A. A Femto voltage amplifier at room temperature was used to amplify the voltage across the device. A gate-voltage $V_{\rm g}$ ranging between $-$80 to 80 V was applied to the Si substrate with a 10 k$\Omega$ series resistance. The magnetic field was applied perpendicular to the $\mathrm{MoS_2}$ plane using a superconducting solenoid. Temperature was measured using a Cernox temperature sensor placed close to the sample.

\section{Theoretical Background}
\label{Theo}

The SI state in 2D granular superconductors and the related SIT can exhibit either Fermionic or a Bosonic character. In the Fermionic case, when the SC order parameter vanishes --- either due to increased disorder or an applied magnetic field --- a transition to a Fermionic insulating state \cite{valles1992electron, hsu1993magnetic, hollen2013collapse} can occur. This insulating phase is also characterized by an activated zero-bias conductance, with the transport governed by quasiparticles (QPs) that resemble unpaired non-interacting electrons. The activation energy, discussed later, does not include the pair breaking energy since the SC order is already destroyed.

The Bosonic SI state, on the other hand, coexists with the intra-grain SC order but without inter-grain phase coherence \cite{baturina2007localized, steiner2008approach, hollen2011cooper}. Here, a 2D granular superconductor is modeled as a random 2D JJ array with a distribution in the inter-grain Josephson coupling energy $E_{\rm J}$ and Coulomb blockade energy $E_{\rm C}=e^2/2C$ with $C$ as the grain capacitance. The SIT occurs when $E_{\rm C}/E_{\rm J}$ exceeds one. Below, we elaborate on the charge BKT physics in the Bosonic SI state that can dictate the conductance behavior at low temperatures. This is followed by phenomenological models that describe role of thermal activation and quantum fluctuations to QP transport.

\subsection{Coulomb Blockade and Charge BKT physics}
\label{sec:Bosonic}

The Bosonic SI state arises in SC grains with very weak Josephson coupling leading to the localization of CPs on the SC grains while the vortices move freely through the region between grains. Several phenomena and observables, which are hallmark of SC state, have corresponding dual in the SI state. For example, the SI state exhibits a critical voltage $V_{\rm C}$ below which a zero current state persists \cite{vinokur2008superinsulator}. This $V_{\rm C}$ is dual to the critical current $I_{\rm C}$ of the SC state below which a zero voltage state persists. Furthermore, the dual of the vortex BKT transition \cite{halperin1979resistive} in a clean 2D SC is the charge BKT transition in the SI state. The vortices, with logarithmic dependence of inter-vortex interaction energy on their separation, have CP solitons as their dual counterparts \cite{delsing1994charge}. The interaction energy between these solitons is proportional to $E_{\rm C}$ and it also exhibits a logarithmic dependence on inter-soliton separation `$r$' in certain limits \cite{mooij1990unbinding,fazio1991charge}. Thus, a charge BKT transition is predicted at a temperature $T_{\rm CBKT} \cong E_{\rm C}/\pi \varepsilon k_{\rm B}$ \cite{mooij1990unbinding,fazio1991charge} with $\varepsilon$ as a non-universal constant of unity order. Below $T_{\rm CBKT}$, the CP soliton and anti-CP soliton remain bound, forming neutral CP dipoles. These bound pairs are localized, resulting in the SI state with ``zero conductance". Above $T_{\rm CBKT}$, thermal fluctuations cause these dipoles to break, resulting in free CP solitons. The conductance in this regime scales with free CP soliton density and is given by \cite{mooij1990unbinding,mironov2018charge},
\begin{equation}
G = G_0 \exp \bigg(\frac{-2b}{\sqrt{(T/T_{\rm CBKT})-1}} \bigg).
\label{eq:ChargeBKT}
\end{equation}
Here, $G_0$ is a pre-factor, and $b$ is a constant of the order of unity.

\begin{figure*}
	\centering
 	\includegraphics[width=6.4in]{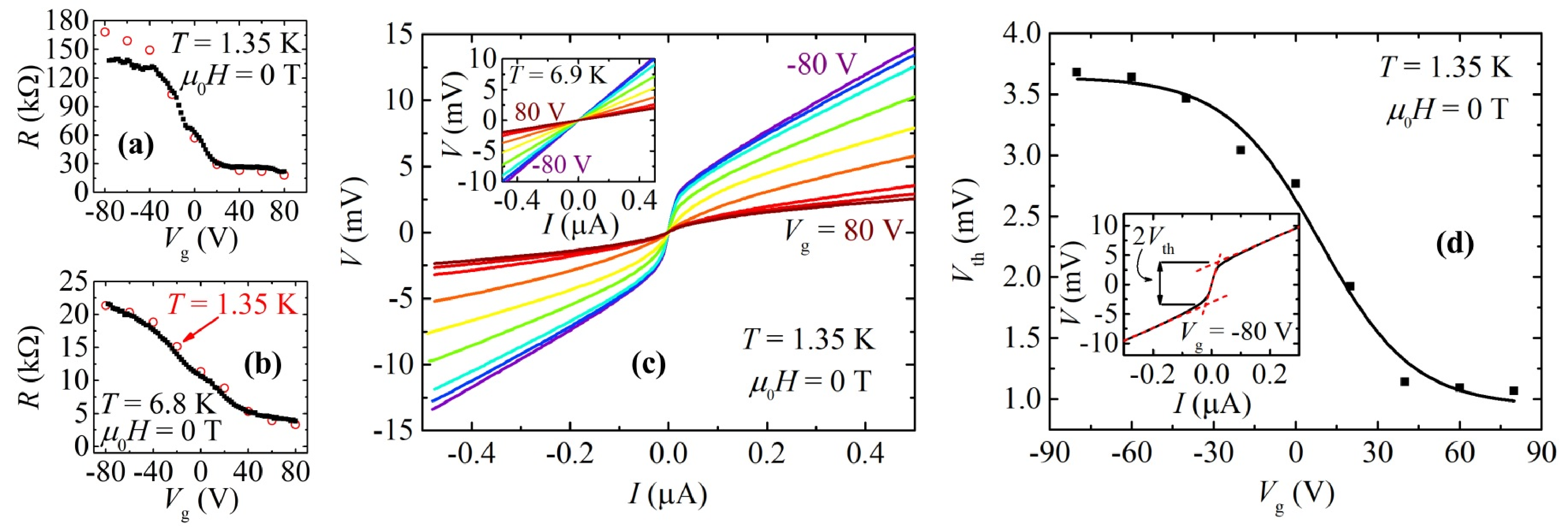}
	\caption{The filled squares in (a) and (b) show the $V_{\rm g}$ dependence of $R$ at 10 nA bias current and at $T =$ 1.34 and 6.8 K, respectively. The open circles in (a) correspond to the zero bias slope ${\rm d}V/{\rm d}I$ while those in (b) correspond to the large bias ${\rm d}V/{\rm d}I$, both at 1.35 K. These slopes are obtained from $V-I$ characteristics shown in (c) and as depicted by red dashed lines in the inset of (d). (c) $V-I$ characteristics at $T =$ 1.35 K for $V_{\rm g}$ values between $-$80 and 80 V and at 20 V $V_{\rm g}$ interval. The inset shows the VICs for the same $V_{\rm g}$ values but at $T =$ 6.9 K. (d) shows the variation of $V_{\rm th}$ with $V_{\rm g}$. The inset defines the $V_{\rm th}$, with the black line showing the VIC at $V_{\rm g}=-80$ V and the red lines marking the zero and large bias behaviors.}
	\label{fig:INS2}
\end{figure*}
The vortex BKT transition in a 2D SC is known to be affected by the presence of vortex pinning. Sharp vortex BKT transitions have been observed in 2D SCs with extremely weak disorder \cite{zhao2020vortex,saito2015metallic,zhao2013evidence} or in SCs with strong but homogeneous disorder \cite{weitzel2023sharpness} at atomic scale. Both these amount to very weak or no pinning of vortices. The vortex-pinning leads to a deviation from the expected BKT behavior at low temperatures \cite{gupta2024gate, benfatto2009broadening} where the vortex pinning can dominate over the inter-vortex interaction. The pinning reduces the vortex mobility, which is beyond the vortex-BKT model. Thus, experimental 2D SC systems with negligible vortex pinning have played a crucial role in establishing vortex-BKT transition. Similar to vortex-BKT model, an ideal SI state with zero free CP solitons is expected below the charge BKT transition temperature $T_{\rm CBKT}$ \cite{mironov2018charge, ikegami2022insulating, katsumoto1994charge}. However, it is not clear how the large inhomogeneity associated with granularity, which is needed for the Coulomb blockade and the SI state, would affect the CP soliton's pinning and the charge BKT physics.

\subsection{Quasiparticle transport across inter-grain barrier}
\label{sec:Fermi}
The suppression of zero-bias conductance below SC $T_{\rm C}$ in a 2D granular superconductor that exhibits an SI state can also be viewed from the inter-grain charge transport perspective. It turns out that a large inter-grain barrier hinders the charge transport between two SC grains much more than that between the normal metal grains. For very small inter-grain coupling, or $E_{\rm J}$, the CP transport gets ruled out and at the same time, the single-electron-like QPs are scarce in the SC grains. The consequence of this is well known in a low transmission tunnel junction between two superconductors whose VIC is dictated by QPs with a negligible conductance below a threshold voltage of $2\Delta/e$. Here $\Delta$ is the BCS gap parameter. Thus, the zero-bias conductance, arising from thermally activated QPs, in such a junction has an activated behavior as it goes like $e^{-2\Delta/k_BT}$.

The coupling between SC grains can be mediated by different media, such as metal, semiconductor or a thin insulator. For mediation by a normal metal, the Cooper-pair transport is dictated by Andreev reflection. In this case, the electron coherence in the metal is important for coherent CP transmission. Thus, if the metal length, between superconductors, exceeds its coherence length, $E_{\rm J}$ becomes negligible. Further, the quality of the interface between the metal and SC plays a crucial role as a poor interface transparency suppresses the Andreev reflection. In the latter case, the thermally activated QPs will dominate the low-bias transport.

In case of a non-degenerate semiconductor mediating the transport between SC grains, the charge transport can only happen through QPs. The Schottky barrier at the semiconductor and SC interface can further hamper the transport. In a moderately n-doped semiconductor, its Fermi level $E_{\rm F}$ (or chemical potential) is below its conduction band minimum $E_{\rm CB}$ and an electron will face a barrier $E_{\rm CB}-E_{\rm F}$ to get transported via the semiconductor. In case of few layer MoS$_2$ as semiconductor, this barrier is tunable by back-gate voltage. The single electron Coulomb charging energy can also add to this barrier in case of low inter-grain transmission but this addition will be same for normal and SC state of the grains. Further, for QP transport between SC grains, the pair breaking energy is another additional energy cost. In case of thin insulating barriers mediating transport via quantum tunneling between grains, the Coulomb charging energy $E_{\rm C}$ can hinder the transport between normal as well as SC grains. Delsing et al. \cite{delsing1994charge} studied an array of tunnel junctions with large $E_{\rm C}$ and negligible $E_{\rm J}$. They argued that, in the normal state, an activation energy of $E_{\rm C}/4$ is required for overcoming the charging barrier for an electron to tunnel between grains. In the SC state, there is an additional energy cost of generating the QPs. Thus, in general we can conclude that the inter-grain charge transport is more hindered in the SC state of the weakly coupled grains than in their normal state.

\subsection{Thermal and quantum fluctuation controlled transport}
\label{sec:ThermalQuantumFermi}

Eventually, the activated conductance with an activation energy barrier $U$ can be described by its temperature dependence as below:
\begin{equation}
G = G_0\:\exp\bigg(-\frac{U}{k_{\rm B} T}\bigg).
\label{eq:ArrheniusEq}
\end{equation}
Here $G_0$ is a constant. In the Pb-MoS$_2$ hybrid system under study, $E_{\rm J}$ is considerably smaller than $k{\rm _B}T$, while the activation barrier $U$ can exceed $k{\rm _B}T$ as well as $E_{\rm J}$. The possible origins of activation barrier were discussed above. According to this expression, the conductance decreases monotonically with cooling and vanishes at zero temperature.

However, as seen later, the conductance in the Pb-MoS$_2$ hybrid system is found to saturate at low temperatures. Such saturation of $G$ has previously been observed in 2D arrays \cite{delsing1994charge} and granular SCs \cite{zhang2022anomalous} and attributed to quantum fluctuations or quantum tunneling of charge carriers between grains, as proposed by Delsing et al. \cite{delsing1994charge}. In this case, the conductance is given by \cite{delsing1994charge, zhang2022anomalous}:
\begin{equation}
G = \bigg ( \frac{1}{R_0} - \frac{1}{R_{\rm S}} \bigg ) \exp \bigg ( -\frac{U}{k_{\rm B}T}\bigg ) + \frac{1}{R_{\rm S}}.
\label{eq:ThermalQuantumEq}
\end{equation}
Here, $R_0$ is associated with the normal-state resistance, and $R_{\rm S}$ is the saturation resistance value.

\section{Results and analysis}
\subsection{Gate dependent transport measurements}
\label{RT_Vg}

The filled squares in Fig. \ref{fig:INS2}(a) show the $V_{\rm g}$ dependence of the resistance of the MoS$_2$-Pb device at $T =$ 1.35 K measured at 10 nA DC bias-current. A monotonic decrease in resistance with increase in $V_{\rm g}$ is attributed to n-doped nature of MoS$_2$ which leads to an increase in its carrier density or to a shift in its Fermi energy towards the conduction band edge. Since the Fermi level lies within the bandgap of MoS$_2$ and near the conduction band edge, low charge carrier density in it leads to poor coupling with a negligible $E_{\rm J}$ between the Pb grains. An $R$ Vs $V_{\rm g}$ at 6.8 K, which is close to the $T_{\rm C}$ of bulk Pb, is shown in Fig. \ref{fig:INS2}(b). Here the resistance values over the studied $V_{\rm g}$ range are substantially smaller as compared to that for $T =$ 1.35 K. Although the $V_{\rm g}$ dependence of the two look similar, the 1.35 K plot exhibits relatively abrupt jump over $V_{\rm g}$ range from $-30$ to 30 V.

\begin{figure*}
	\centering
 	\includegraphics[width=6.2in]{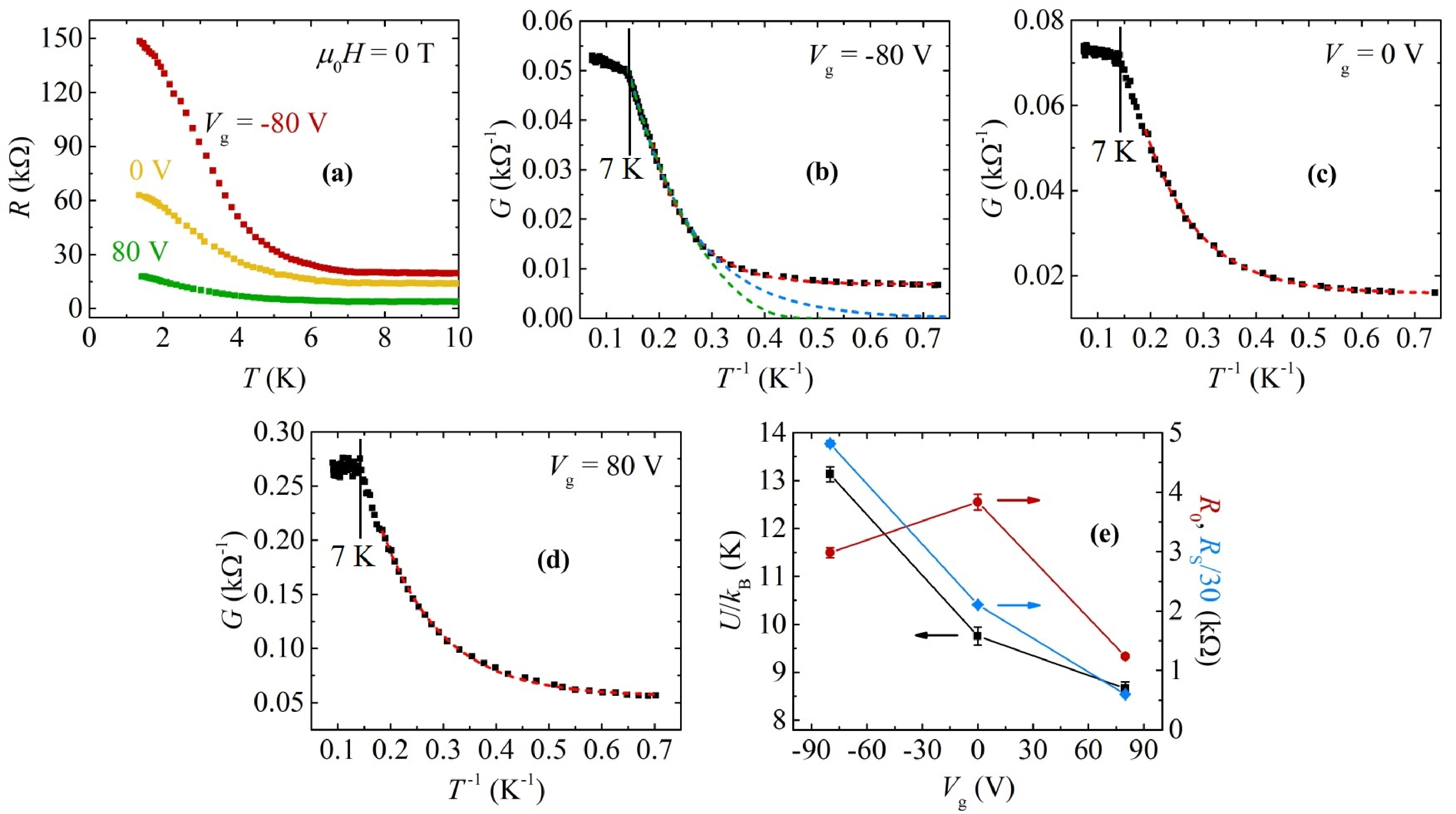}
	\caption{(a) shows zero field $R$ (at 10 nA DC current) Vs $T$ at three different $V_{\rm g}$ values. The filled squares in (b), (c) and (d) show the conductance $G = R^{-1}$ as a function of $T^{-1}$ for $V_{\rm g}= -$80, 0, and 80 V, respectively. The red lines in (b), (c) and (d) are fits to equation (\ref{eq:ThermalQuantumEq}) and the green and blue lines in (b) are fits to equations (\ref{eq:ChargeBKT}) and (\ref{eq:ArrheniusEq}), respectively. (e) shows the fitting parameters $U$, $R_0$, and $R_{\rm S}$ obtained from the fits to equation (\ref{eq:ThermalQuantumEq}) and as a function of $V_{\rm g}$.}
	\label{fig:INS3}
\end{figure*}
Figure \ref{fig:INS2}(c) shows the zero magnetic-field $V-I$ characteristics of the sample for $V_{\rm g}$ values from $-$80 to +80 V at $T = 1.35$ K. The VICs exhibit nonlinearity, with the slope ${\rm d}V/{\rm d}I$ changing from a large value at small bias currents to a lower value at large currents. This slope change occurs at a critical or threshold voltage $V_{\rm th}$, which is defined in the inset of Fig. \ref{fig:INS2}(d) as the voltage at which the linearly extrapolated low-bias and high-bias VIC portions intersect. $V_{\rm th}$ decreases monotonically with an increase in $V_{\rm g}$, as displayed in Fig. \ref{fig:INS2}(d). Such large slope ${\rm d}V/{\rm d}I$ at zero bias-current and with a critical or threshold voltage $V_{\rm th}$ can be attributed to the SI behavior \cite{vinokur2008superinsulator} discussed earlier. Note that even below $V_{\rm th}$, the conductance, or the current, does not vanish. This can be attributed to thermal activation and quantum fluctuations.

The $V_{\rm g}$ dependence of $V_{\rm th}$ deduced from Fig. \ref{fig:INS2}(d) is similar to the $R$ dependence of $V_{\rm g}$ at $T =$ 1.35 K of Fig. \ref{fig:INS2}(a) as both show a relatively abrupt drop in a $V_{\rm g}$ range from $-$30 to +30 V. Fig. \ref{fig:INS2}(c) inset shows the VICs at 6.9 K, which are linear with a $V_{\rm g}$ dependent slope ${\rm d}V/{\rm d}I$ consistent with the measured $R$ Vs $V_{\rm g}$, see Fig. \ref{fig:INS2}(a). The $V_{\rm g}$ dependent large-bias (\emph{i.e.} well above $V_{\rm th}$) slope at 1.35 K, see the red circles in Fig. \ref{fig:INS2}(b), is found to be the same as the $R$ Vs $V_{\rm g}$ at 6.8 K. This is noteworthy and implies that above $V_{\rm th}$ the VICs exhibit the normal-state resistance of above $T_{\rm C}$. This also implies a nearly temperature independent behavior of the normal-resistance even below $T_{\rm C}$.

Figure \ref{fig:INS3} (a) shows the zero field resistance (at 10 nA current) of the device as a function of temperature below 10 K and at $V_{\rm g} = -$80, 0, and 80 V. The resistance change between 10 and 7 K is extremely small. However, below about 7 K, \emph{i.e.} $T_{\rm C}$ of bulk Pb, a rise in $R$ is observed with decreasing $T$. This is more apparent in the conductance $G (= 1/R)$ Vs $T^{-1}$ activation plots in Figs. \ref{fig:INS3}(b-d) where a clear kink in $G$ is seen at 7 K. We attribute this sharp drop in conductance occurring at the SC transition temperature of Pb grains, \emph{i.e.} 7 K, to the onset of Bosonic SI state. Also, there is a small slope, in these activation plots, at large negative $V_{\rm g}$ values, for $T>7$ K. This indicates a slight insulating behavior above 7 K at large negative $V_{\rm g}$ values which gradually disappears as $V_{\rm g}$ is increased. The slight insulating nature seems to get converted into a much stronger insulator as one enters the SC state.

We have attempted to fit the $G$ versus $T^{-1}$ plot in Fig. \ref{fig:INS3}(b) with charge-BKT transition, \emph{i.e.} Eq. (\ref{eq:ChargeBKT}), as shown by the green dashed line. The equation fits well within a temperature range starting from $T_{\rm C}$ and down to a certain temperature; below which, the fit deviates as the measured $G$ saturates to a finite value at low temperatures. This deviation could arise from the finite size of the sample, as suggested in the literature \cite{mironov2018charge}, or from sample inhomogeneity, which leads to a distribution of $E_{\rm C}$ values. Alternatively, it could be due to some limitations of the model itself that may include the pinning of CP solitons, as discussed earlier.

\begin{figure*}
	\centering
 	\includegraphics[width=6.5in]{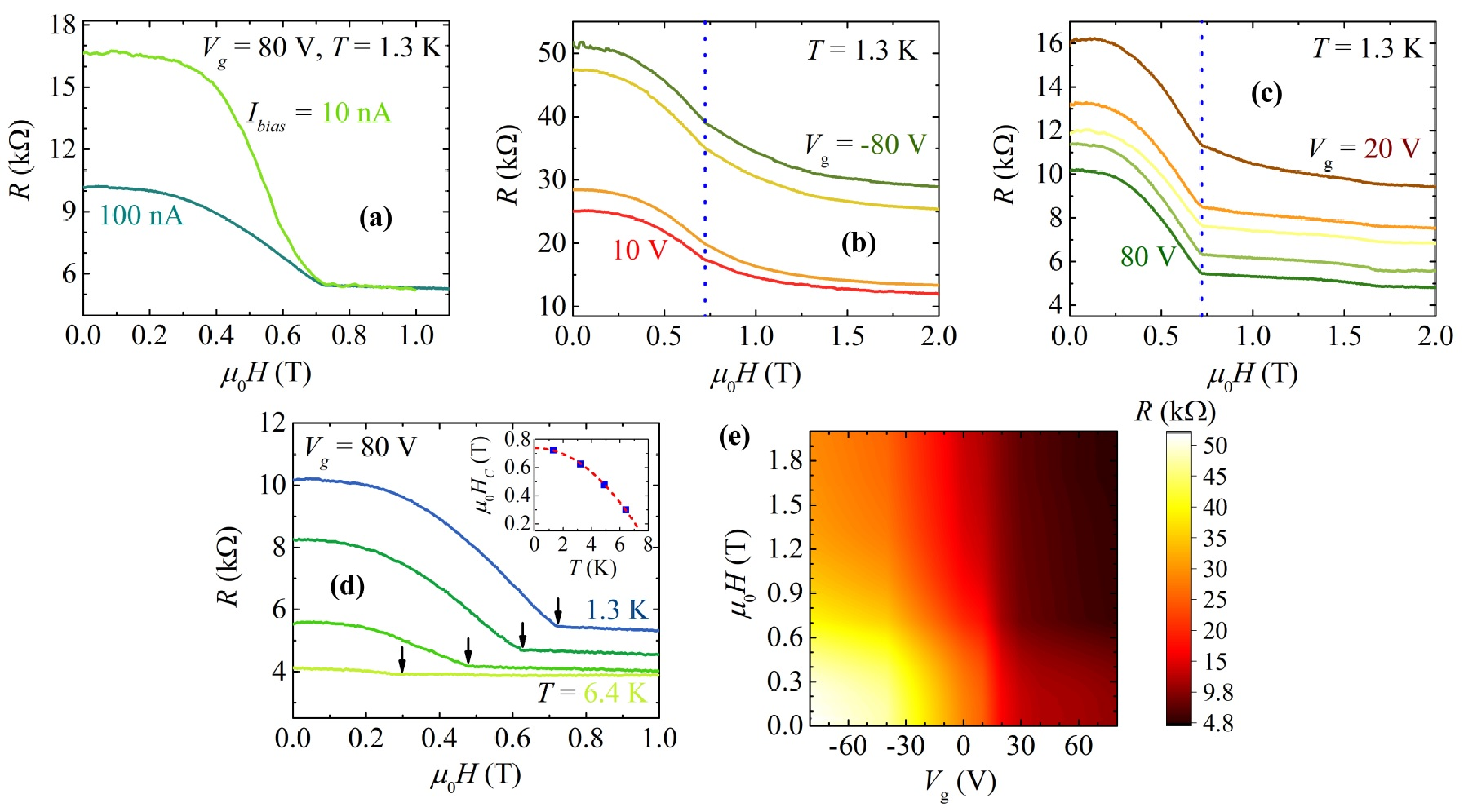}
	\caption{(a) Magnetic field $\mu_0 H$ dependence of resistance for bias currents 10 and 100 nA at $V_{\rm g} =$ 80 V and $T =$ 1.3 K. (b) and (c) show the field dependence of $R$ (at 100 nA)  at $T =$ 1.3 K and at $V_{\rm g} = -$80, $-$40, 0, 10 V and at $V_{\rm g} =$ 20, 30, 40, 60 and 80 V, respectively. The blue dotted lines in (b) and (c) at $\mu_0 H =$ 0.72 T marks the critical field $\mu_0 H_{\rm C}$. (d) Variation of $R$ (at 100 nA) with $\mu_0 H$ at $V_{\rm g} =$ 80 V for $T =$ 1.3, 3.2, 4.9 and 6.4 K. The black arrows mark $\mu_0 H_{\rm C}$ for different temperatures. The inset shows the plot of this $\mu_0 H_{\rm C}$ with $T$ with the red line as the fit to equation (\ref{BT}). (e) 2D color map of $R$ (at 100 nA) in the $V_{\rm g} - \mu_0 H$ plane at $T =$ 1.3 K.}
	\label{fig:INS4}
\end{figure*}
Taking the inter-grain barriers into consideration, we have attempted to fit the conductance using the thermally activated conductance given by Eq. \ref{eq:ArrheniusEq}. This fit, shown by the blue dashed line in Fig. \ref{fig:INS3}(b), also deviates below nearly the same temperature where the fit to charge-BKT equation begins to fail. Finally, the Eq. \ref{eq:ThermalQuantumEq}, which includes contributions from both the thermal activation and quantum tunneling processes fits well to the data over the whole studied temperature range below 7K, see the red dashed lines in Figs. \ref{fig:INS3}(b-d). Fig. \ref{fig:INS3}(e) plots the fitting parameters $U$, $R_0$, and $R_{\rm S}$ extracted from the fits in Figs. \ref{fig:INS3}(b-d), as a function of $V_{\rm g}$. The activation energy $U$ decreases with increasing $V_{\rm g}$, as does $R_{\rm S}$, while $R_0$ exhibits a non-monotonic dependence on $V_{\rm g}$. $U$ is related to the inter-grain barrier with contributions from various factors discussed in section \ref{sec:Fermi}.

\subsection{Magnetic field dependent transport}
\label{RT_B}

Figure \ref{fig:INS4}(a) shows the resistance $R$ as a function of the applied magnetic field $H$ at $V_{\rm g} =$ 80 V and $T =$ 1.3 K, for two bias currents, 10 and 100 nA. $R$ at low fields measured with 10 nA current is nearly twice of that measured with 100 nA. This is due to the non-linearity of the VICs, see Fig. \ref{fig:INS2}(c). At 10 nA, the sample voltage is well below $V_{\rm th}$ while at 100 nA it is close to $V_{\rm th}$. We also see that $R$ has little variation at low fields of up to about 0.4 T and there is a sharp decline beyond this. Furthermore, Fig. \ref{fig:INS4}(a) shows that $R(\mu_0 H)$ curves at two bias currents coincide above $\mu_0 H = 0.72$ T. This happens as the VICs become linear above 0.72 T field. We attribute this behavior to magnetic field-induced pair breaking \cite{hsu1993magnetic, kim1992superconductivity} in the Pb grains. Thus, 0.72 T is the depairing or the critical field $\mu_0 H_{\rm C}$ above which the superconductivity in the Pb islands is lost. Note that critical field of bulk Pb is 0.08 T but it gets enhanced by several orders for small grains depending on their size \cite{Li-nano-Pb}. The disappearance of VIC nonlinearity and vanishing of $V_{\rm th}$ above a critical field, presumably same as that of SC grains, is again consistent with the SI state \cite{vinokur2008superinsulator}.
\begin{figure}
	\centering
  	\includegraphics[width=3.2in]{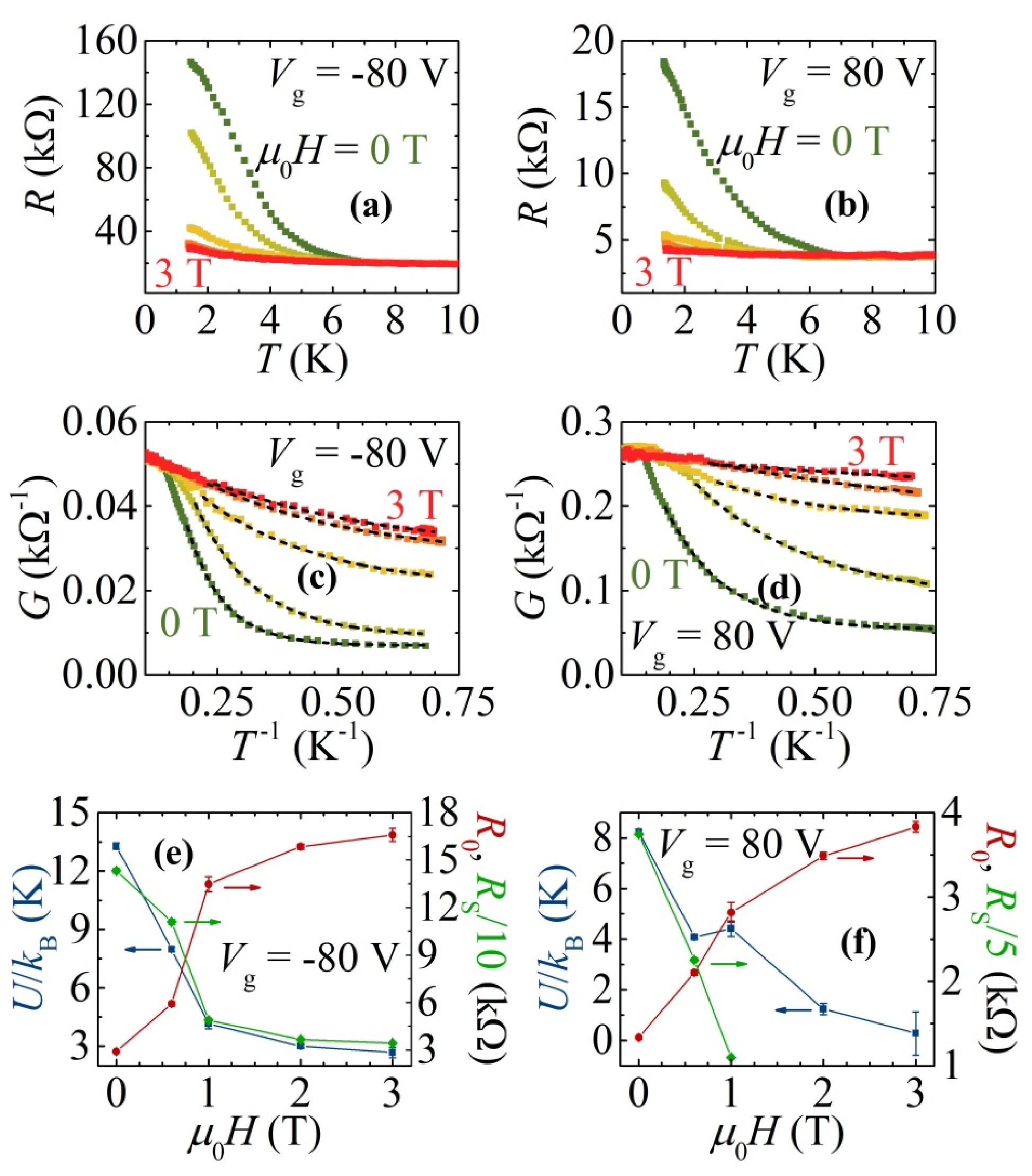}
	\caption{(a) and (b) show $R$ (at 10 nA) as a function of $T$, for $\mu_0 H =$ 0, 0.58, 1, 2 and 3 T for $V_{\rm g} = -$80 and 80 V, respectively. (c) and (d) show the corresponding plots of conductance (colored squares) as a function of $T^{-1}$ for $V_{\rm g} = -$80 V and 80 V, respectively. The black dashed lines are the fits to equation (\ref{eq:ArrheniusEq}). (e) and (f) show the fitting parameters $U$, $R_0$ and, $R_{\rm S}$, see equation (\ref{eq:ThermalQuantumEq}), as a function of $\mu_0 H$ for $V_{\rm g} = -$80 V and 80 V, respectively.}
	\label{fig:INS5}
\end{figure}

Figures \ref{fig:INS4}(b,c) show $R(\mu_0 H)$ curves taken at 100 nA for various $V_{\rm g}$ values at $T = 1.3$ K. Although for this bias current, $V$ is close to $V_{\rm th}$ but the characteristic features of SI state remain discernible. The kink-like features in Figs. \ref{fig:INS4}(b,c) are marked by the blue dotted lines, marking the depairing field. Notably, this field is independent of $V_{\rm g}$. This can be expected as intra-grain superconductivity of Pb is unaffected by $V_{\rm g}$. Fig. \ref{fig:INS4}(d) plots $R(\mu_0 H)$ for $V_{\rm g} = 80$ V at different temperatures where $\mu_0 H_{\rm C}$, marked by the black arrows, is clearly seen to be temperature dependent. $\mu_0 H_{\rm C}$ and decreases with increasing $T$ as seen in the inset of Fig. \ref{fig:INS4}(d). The red dashed line in this plot is the fit to equation \cite{barber2006negative}:
\begin{equation}
H_{\rm C}(T) = H_{\rm C}(0) \bigg [ 1- \bigg (\frac{T}{T_{\rm C}}\bigg)^2\: \bigg ],
\label{BT}
\end{equation}
with $\mu_0 H_{\rm C}(0)=0.738$ T. Fig. \ref{fig:INS4}(e) shows a 2D color map of $R$ in the $V_{\rm g}-\mu_0 H$ plane at $T =$ 1.3 K. The bright white region outlined by the yellow boundary at low fields and large negative $V_{\rm g}$ corresponds to the SI regime. 

Figures \ref{fig:INS5}(a,b) show the $R(T)$ curves at different magnetic fields for $V_{\rm g} = -$80 and 80 V, respectively. Here, $R$ decreases with increasing $\mu_0 H$ at all temperatures below $T_{\rm C}$ and it is nearly fixed above $T_{\rm C}$. The decrease is more pronounced at low temperatures. This negative magneto-resistance behavior can be attributed to increased QP density in Pb grains with increasing field \cite{gantmakher2010superconductor, gerber1997insulator}. Note that this behavior is opposite to that of superconductors which show a positive magneto-resistance. Further, the resistance at $\mu_0 H =$ 1 T, which is above $\mu_0 H_{\rm C}$, exhibits an upturn at low temperatures and for both extreme gate voltages and at $T =$ 1.3 K, $R$ remains higher than that at 7 K. Such an insulator-like behavior above the depairing field and at low temperatures is consistent with the presence of a small intergrain transport barrier even for the normal electrons. Another related observation is the resistance tail in Figs. \ref{fig:INS4}(a-d) that lasts well above 0.72 T and is more pronounced at negative $V_{\rm g}$ values. Both these observations could indicate a crossover from a Bosonic to a Fermionic insulator \cite{gerber1997insulator} at the depairing field. It is also possible that the SC correlations persist well above the depairing field.

Figures \ref{fig:INS5}(c,d) show $G$ Vs $T^{-1}$ for $V_{\rm g} = -$80 and 80 V, respectively, and at different magnetic fields together with the fits to Eq. \ref{eq:ThermalQuantumEq}. The obtained fitting parameters $U$, $R_0$, and $R_{\rm S}$ are plotted as functions of $\mu_0 H$ in Figs. \ref{fig:INS5} (e,f) for the two gate voltages. Expectedly, $U$ decreases monotonically with increasing $\mu_0 H$ due to reduction of the inter-grain barrier with field induced de-pairing \cite{hsu1993magnetic}. This decrease in $U$ is much steeper for fields below 1 T than above it. Additionally, $R_{\rm S}$ decreases with increasing $\mu_0 H$, while $R_0$ increases, with the latter approaching the normal-state resistance at high fields.

\section{Discussion and conclusions}
The above results on the MoS$_2$-Pb device are, at least qualitatively, well described by the SI state below the $T_{\rm C}$ of Pb grains with a gate tunable inter-grain coupling providing a handle on the SI state. Few observations may need further attention; these are: 1) failure of the charge-BKT fit, 2) failure of the thermal activation fit and, 3) the correlated persistence of resistance tail and insulating behavior well above the depairing field. The first two points relate to the conductance saturation with cooling as opposed to the expected continuous decline. As mentioned earlier, it is not clear how a large disorder, essential for the SI state existence, will affect the charge-BKT physics. Some deviation from the ideal charge-BKT behavior in terms of broadening of the transition can be anticipated from the effect of disorder on vortex-BKT physics. The failure of thermal activation and role of quantum fluctuations has been invoked in systems exhibiting SI behavior \cite{delsing1994charge}. Here, it works surprisingly well and over a wide range of $V_{\rm g}$ and magnetic field. The third point on resistance tail persisting up to large fields, particularly for insulating state at large negative $V_{\rm g}$. The insulating behavior in the normal state of grains looks normal but it is not fully clear how it can lead to a resistance tail up to fields beyond the depairing field. A crossover from Bosonic to Fermionic SI is a possibility that needs to be more investigations.

Further, in the studied device the exact origin of the gate tunable inter-grain barrier is not very clear. As pointed out in section \ref{sec:Fermi}, this barrier may have several contributors including Coulomb charging energy, BCS gap $\Delta$, Schottky barrier and the barrier $E_{\rm CB} - E_{\rm F}$ in MoS$_2$. From the gate tunability of conductance, it is clear that either or both of the last two must be playing some role together with others. An access to degenerately doped regime of MoS$_2$ may be of help and further interest as MoS$_2$ itself may become a superconductor in this regime \cite{Saito-MoS2-supercond}.

In conclusion, the temperature dependent electrical transport study on MoS$_2$-Pb device at different gate-voltages and magnetic fields unravel a super-insulating phase below the critical temperature and critical field of the Pb grains. The underlying physics can be understood, at least naively, from the enhanced hindrance in inter-grain charge transport when the Pb grains are superconducting. The conductance saturation at low temperature is surprisingly well modeled by a combination of thermal activation and quantum fluctuations. Further, a crossover from Bosonic to a Fermionic insulator may be happening when the inter-grain barriers become stronger; however this needs more conclusive experiments.

\section*{ACKNOWLEDGMENTS}
We acknowledge SERB-DST of the Government of India and IIT Kanpur for financial support.


\begin{thebibliography}:
\bibitem{sondhi1997continuous} Sondhi S L, Girvin S, Carini J and Shahar D 1997 {\it Rev. Mod. Phys.} {\bf 69} 315

\bibitem{fisher1990quantum} Fisher M P A 1990 {\it Phys. Rev. Lett.} {\bf 65} 923

\bibitem{tsen2016nature} Tsen A, Hunt B, Kim Y, Yuan Z, Jia S, Cava R, Hone J, Kim P, Dean C and Pasupathy A 2016 {\it Nat. Phys.} {\bf 12} 208--212

\bibitem{sharma20182d} Sharma C H, Surendran A P, Varma S S and Thalakulam M 2018 {\it Commun. Phys.} {\bf 1} 90

\bibitem{allain2012electrical} Allain A, Han Z and Bouchiat V 2012 {\it Nat. Mater.} {\bf 11} 590--4

\bibitem{breznay2017superconductor} Breznay N P, Tendulkar M, Zhang L, Lee S C and
Kapitulnik A 2017 {\it Phys. Rev. B} {\bf 96} 134522

\bibitem{xing2015quantum} Xing Y, Zhang H M, Fu H L, Liu H, Sun Y, Peng J P,
Wang F, Lin X, Ma X C, Xue Q K, Wang J and Xie X C 2015 {\it Science} {\bf 350} 542--5

\bibitem{chen2024quantum} Chen F, Liu Y, Guo W, Wang T, Tian Z, Zhang M, Xue Z, Mu G, Zhang X and Di Z 2024 {\it Nano Lett.} {\bf 24} 2444--50

\bibitem{nguyen2009observation} Nguyen H, Hollen S M, Stewart Jr M, Shainline J, Yin A, Xu J and Valles Jr J M 2009 {\it Phys. Rev. Lett.} {\bf 103} 157001

\bibitem{hollen2014fate} Hollen S M, Fernandes G, Xu J and Valles Jr J 2014 {\it Phys. Rev. B} {\bf 90} 140506

\bibitem{baturina2007localized} Baturina T, Mironov A Y, Vinokur V, Baklanov M and Strunk C 2007 {\it Phys. Rev. Lett.} {\bf 99} 257003

\bibitem{stewart2007superconducting} Stewart Jr M, Yin A, Xu J and Valles Jr J M 2007 {\it Science} {\bf 318} 1273--5

\bibitem{khan2000superconductor} Khan S, Pedersen E, Kain B, Jordan A and Barber Jr R
2000 {\it Phys. Rev. B} {\bf 61} 5909

\bibitem{hen2021superconductor} Hen B, Zhang X, Shelukhin V, Kapitulnik A and Palevski
A 2021 {\it PNAS} {\bf 118} e2015970118

\bibitem{van1996quantum} van der Zant H S, Elion W J, Geerligs L J and Mooij J 1996 {\it Phys. Rev. B} {\bf 54} 10081

\bibitem{han2014collapse} Han Z, Allain A, Arjmandi-Tash H, Tikhonov K,
Feigel'man M, Sac{\'e}p{\'e} B and Bouchiat V 2014 {\it Nat. Phys.} {\bf 10} 380--6

\bibitem{vinokur2008superinsulator} Vinokur V M, Baturina T I, Fistul M V, Mironov A Y, Baklanov M R and Strunk C 2008 {\it Nature} {\bf 452} 613--5

\bibitem{fistul2008collective} Fistul M, Vinokur V and Baturina T 2008 {\it Phys. Rev. Lett.} {\bf 100} 086805

\bibitem{baturina2013superinsulator} Baturina T I and Vinokur V M 2013 {\it Ann. Phys.} {\bf 331} 236--57

\bibitem{steiner2008approach} Steiner M A, Breznay N P and Kapitulnik A 2008 {\it Phys. Rev. B} {\bf 77} 212501

\bibitem{hollen2011cooper} Hollen S M, Nguyen H Q, Rudisaile E, Stewart Jr. M D, Shainline J, Xu J M and Valles Jr J M 2011 {\it Phys. Rev. B} {\bf 84} 064528

\bibitem{valles1992electron} Valles Jr J M, Dynes R C and Garno J P 1992 {\it Phys. Rev. Lett.} {\bf 69} 3567

\bibitem{hsu1993magnetic} Hsu S-Y and Valles Jr J M 1993 {\it Phys. Rev. B} {\bf 48} 4164

\bibitem{hollen2013collapse} Hollen S M, Fernandes G E, Xu J M and Valles Jr J M 2013 {\it Phys. Rev. B} {\bf 87} 054512
\bibitem{sambandamurthy2004superconductivity} Sambandamurthy G, Engel L, Johansson A and Shahar D 2004 {\it Phys. Rev. Lett.} {\bf 92} 107005

\bibitem{sambandamurthy2005experimental} Sambandamurthy G, Engel L, Johansson A, Peled E and Shahar D 2005 {\it Phys. Rev. Lett.} {\bf 94} 017003

\bibitem{beloborodov2007granular} Beloborodov I S, Lopatin A V, Vinokur V M and Efetov K B 2007 {\it Rev. Mod. Phys.} {\bf 79} 469--518

\bibitem{syzranov2009dc} Syzranov S V, Efetov K B and Altshuler B L 2009 {\it Phys. Rev. Lett.} {\bf 103} 127001

\bibitem{radisavljevic2011single} Radisavljevic B, Radenovic A, Brivio J, Giacometti V and Kis A 2011 {\it Nat. Nanotechnol.} {\bf 6} 147--50

\bibitem{yoon2011good} Yoon Y, Ganapathi K and Salahuddin S 2011 {\it Nano Lett.} {\bf
11} 3768--73

\bibitem{castellanos2014deterministic} Castellanos-Gomez A, Buscema M, Molenaar R, Singh V, Janssen L, Van Der Zant H S and Steele G A 2014 {\it 2D Mater.} {bf 1} 011002

\bibitem{chakraborty2012symmetry} Chakraborty B, Bera A, Muthu D, Bhowmick S, Waghmare
U V and Sood A 2012 {\it Phys. Rev. B} {\bf 85} 161403

\bibitem{lee2010anomalous} Lee C, Yan H, Brus L E, Heinz T F, Hone J and Ryu S
2010 {\it ACS Nano} {\bf 4} 2695--700

\bibitem{halperin1979resistive} Halperin B and Nelson D R 1979 {\it J. Low Temp. Phys.} {\bf 36} 599--616

\bibitem{delsing1994charge} Delsing P, Chen C, Haviland D, Harada Y and Claeson T 1994 {\it Phys. Rev. B} {\bf 50} 3959

\bibitem{mooij1990unbinding} Mooij J, Van Wees B, Geerligs L, Peters M, Fazio R and Sch{\"o}n G 1990 {\it Phys. Rev. Lett.} {\bf 65} 645

\bibitem{fazio1991charge} Fazio R and Sch{\"o}n G 1991 {\it Phys. Rev. B} {\bf 43} 5307

\bibitem{mironov2018charge} Mironov A Y, Silevitch D M, Proslier T, Postolova S V, Burdastyh M V, Gutakovskii A K, Rosenbaum T F, Vinokur V V and Baturina T I 2018 {\it Sci. Rep.} {\bf 8} 4082

\bibitem{zhao2020vortex} Zhao W L, Zhu X, He Z H, Gao K H and Li Z Q 2020 {\it Supercond. Sci. Technol.} {\bf 33} 105010

\bibitem{saito2015metallic} Saito Y, Kasahara Y, Ye J, Iwasa Y and Nojima T 2015 {\it Science} {\bf 350} 409--13

\bibitem{zhao2013evidence} Zhao W, Wang Q, Liu M, Zhang W, Wang Y, Chen M, Guo Y, He K, Chen X, Wang Y, Wang J, Xie X, Niu Q, Wang L, Ma X, Jain J K, Chan M H W and Xue Q K 2013 {\it Solid State Commun.} {\bf 165} 59--63

\bibitem{weitzel2023sharpness} Weitzel A, Pfaffinger L, Maccari I, Kronfeldner K, Huber T, Fuchs L, Mallord J, Linzen S, Il’ichev E, Paradiso N and Strunk C 2023 {\it Phys. Rev. Lett.} {\bf 131} 186002
\bibitem{gupta2024gate} Gupta S, Jana S P, Pervin R and Gupta A K 2024 {\it Phys. Rev. B} {\bf 110} 024506

\bibitem{benfatto2009broadening} Benfatto L, Castellani C and Giamarchi T 2009 {\it Phys. Rev. B} {\bf 80} 214506

\bibitem{ikegami2022insulating} Ikegami H and Nakamura Y 2022 {\it Phys. Rev. B} {\bf 106} 184511

\bibitem{katsumoto1994charge} Katsumoto S and Kobayashi S 1994 {\it Physica B: Condens. Matter} {\bf 194} 1123--4



\bibitem{zhang2022anomalous} Zhang X, Palevski A and Kapitulnik A 2022 {\it PNAS} {\bf 119} e2202496119
\bibitem{kim1992superconductivity} Kim J J, Kim J and Lee H J 1992 {\it Phys. Rev. B} {\bf 46} 11709

\bibitem{Li-nano-Pb} W.-H. Li, C. C. Yang, F. C. Tsao, and K. C. Lee, {\it Phys. Rev. B} {\ bf 68}, 184507 (2003).

\bibitem{barber2006negative} Barber Jr R P, Hsu S Y, Valles Jr J M, Dynes R C and Glover III R E 2006 {\it Phys. Rev. B Condens. Matter} {\bf 73} 134516

\bibitem{gantmakher2010superconductor} Gantmakher V F and Dolgopolov V T 2010 {\it Phys.–Usp.} {\bf 53} 1--57

\bibitem{gerber1997insulator} Gerber A, Milner A, Deutscher G, Karpovsky M and Gladkikh A 1997 {\it Phys. Rev. Lett.} {\bf 78} 4277


\bibitem{Saito-MoS2-supercond} Y. Saito, Y. M. Itahashi, T. Nojima, and Y. Iwasa, Phys. Rev. Mater. 4, 074003 (2020)

\end{thebibliography}
\end{document}